\documentclass[10pt,conference,dvipsnames]{IEEEtran}

\usepackage{graphicx}
\graphicspath{{assets/img/}, {assets/CMYK/}}
\DeclareGraphicsExtensions{.eps,.png,.pdf,}
\ifCLASSOPTIONcompsoc
    \usepackage[caption=false, font=normalsize, labelfont=sf, textfont=sf]{subfig}
\else
\usepackage[caption=false, font=footnotesize]{subfig}
\fi

\usepackage{hyperref}

\usepackage{comment}
\usepackage{todonotes}

\usepackage{amsmath}
\usepackage{amssymb}

\usepackage{balance}

\usepackage{multirow}
\newcounter{cleansubsubcntr}

\usepackage[normalem]{ulem}
\newcommand{\new}[1]{#1}
\newcommand{\newp}[1]{#1}

\newcommand{\del}[1]{}
\newcommand{\delp}[1]{}

\begin{document}

\pdfoutput=1
\title{Sharing without Showing: Secure Cloud Analytics with Trusted Execution Environments}
\author{Marcus Birgersson \\ KTH Royal Institute of Technology \\  marbir@kth.se \and
Cyrille Artho \\ KTH Royal Institute of Technology \\  artho@kth.se \and
Musard Balliu \\ KTH Royal Institute of Technology \\ musard@kth.se \and
}

\maketitle

\def\rqa{How can we design and implement a system that fulfills our requirements?}
\def\rqb{What security and integrity guarantees can we give?}
\def\rqc{What is the overhead of such a system?}



\begin{abstract}
  Many applications benefit from computations over the data of multiple users while preserving confidentiality. 
We present a solution where multiple mutually distrusting users' data can be aggregated with an acceptable overhead, while allowing users to be
added to the system at any time without re-encrypting data. 
Our solution to this problem is to use a Trusted Execution Environment
(Intel SGX) for the computation, while the confidential data
is encrypted with the data owner's key and can be stored anywhere, without trust in the service provider.
We do not require the user to be online during the computation phase and do not require a trusted party to store data in plain text. Still, the computation can only
be carried out if the data owner explicitly has given permission.

Experiments using common functions such as the sum, least square fit, histogram, and SVM classification, exhibit an average overhead of $1.6 \times$.
In addition to these performance experiments, we present a use case for computing the distributions of taxis in a city without revealing the position of any other taxi to the other parties.
\end{abstract}

\begin{IEEEkeywords}
  Multi-party computation,
  Trusted execution platform,
  SGX,
  Confidential computation
\end{IEEEkeywords}

\section{Introduction}
Modern applications often rely on cloud-based infrastructures for storage provisioning and on-demand computing, for example, to carry out data analytics. These advantages come at the price of
trusting the cloud providers with potentially sensitive data, thus increasing the risks for the end users.  

Several studies motivate the need for secure systems that protect sensitive data by design and \del{have the ability to}use this data securely in arbitrary computations~\cite{notConnectingSmartDev,pverifier, healthcareDataBreach}.

Traditional privacy-preserving technologies struggle to mitigate these issues~\cite{fheSurvey,homEncSpringer}. 
Symmetric and asymmetric encryption require encrypted data to be first decrypted to perform any computations, thus making sensitive data accessible to the hosting infrastructure. Homomorphic encryption schemes offer solid security guarantees to compute on sensitive data without revealing any information about the underlying data.
Unfortunately, these solutions come with drawbacks in terms of computation overhead in space and computation time~\cite{homomorphicPerformance}, which can severely affect the application's performance. The problem is further exacerbated in multi-user settings, where computations over data of mutually distrusting users usually require complex secret sharing schemes and users that are active during the computation~\cite{mpcMachineLearning}. The overhead of homomorphic encryption is a factor between tens of thousands and
millions~\cite{homomorphicPerformance}.
Trusted execution environments (TEEs) have been used to confidentially train a model based on multiple users' data~\cite{MlTrainingTee,privPresMlTee,machineLearningTEE}. In those cases, the TEE has mainly been used to protect the \emph{model,} rather than the privacy of the users' \emph{data.} In addition, they either require the user to be online and active during the computation phase or put their trust in a central service provider.

In this paper, we set out to study the above-mentioned challenges by means of confidential computing~\cite{DBLP:journals/iacr/CostanD16}.
We leverage the security properties of confidential computing to design, implement, and evaluate a system that allows us to perform secure computations over sensitive data in a multi-user setting.

Our paper contributes a design that leverages TEEs in an IoT setting with dynamic user participation, i.\,e., users can be added to the computation at any time without any communication with already participating users.
\new{Though our system is in principle more general, it is especially useful for IoT applications. Here, we interpret IoT in a very broad sense: Any device or service that collects data and transports it to a cloud backed is considered part of the IoT ecosystem~\cite{DBLP:journals/ieeesp/BalliuBS19}. IoT devices can be low-powered and not always available from the outside network; hence, users often rely on a secure storage to collect data and, at a later time,  use this data for analytics purposes. Data collection and sharing over a longer period of time requires to accommodate (add or remove) users and this should be done without exchanging new keys with existing users, or  changing the encryption for already collected data. This in an important aspect for long-lived computations in a multi-user setting, since initiating a key-exchange between thousands of users both requires extra resources and that each user is available online.}


Based on the above, we formulate the following research questions:
\begin{itemize}
\item \textbf{RQ1:} \rqa  \
We propose a cloud-based architecture that uses TEEs to perform secure computations over sensitive data from multiple users (Section~\ref{sec:architecture}) and provide a prototype implementation based on Open Enclave (Section~\ref{sec:implementation}). 
\item \textbf{RQ2:} \rqb \
We discuss the security guarantees of our architecture by defining the \new{threat model} and security properties  (Section~\ref{sec:security}).
\item \textbf{RQ3:} \rqc \
We use off-the-shelf software to implement several use cases with acceptable overhead.
The system allows for arbitrary computations of data between mutually distrusting users without any complex secret-sharing schema (Section~\ref{sec:evaluation}).
\end{itemize}

We make the following contributions in this paper:
\begin{enumerate}
  \item We propose a system for confidential computation on \emph{multiple mutually distrusting users} data.
  \item We do not require the users to be online during the computation, only in the initial setup step.
  \item Users can be added at any time without having to change any encryption. 
   \item We implement and evaluate this system over a variety of use cases. 
\end{enumerate}

\section{Background and Motivation}
\label{sec:problem}
\label{sec:background}
Our main concern is the confidentiality of the data in a setting of mutually distrusting users. This implies that unauthorized access to data should be prevented. Moreover, the system should be able to release (declassify) aggregates of sensitive data whenever all users agree upon such a release. 
Our secondary concern is the integrity of data: It should be possible to verify that a computation is carried out as expected, both by the users and by any third party that expects the result.
To realize such a secure multi-user computation system, we use a so-called \emph{Trusted Execution Environment} (TEE)~\cite{teedef}.

\subsection{IoT data privacy}
A lot of smart devices, such as security cameras, baby monitors, or camera doorbells, collect very sensitive data. They store their data online, which makes them vulnerable to data breaches~\cite{iotSecBreach4}.
To mitigate such risks, one can encrypt the data before uploading it to the cloud\del{, using a key that only the user has access to}. The main reason for not doing so is to be able to perform analysis and aggregations on the data in the cloud. In this paper, we will present a solution where data can be encrypted at rest but can still be analyzed without the risk of leaking any confidential data.

This work builds on the work by Birgersson et al.~\cite{secAwareIot}. The authors present an architecture using mandatory access control to prevent the leakage of aggregated data to unauthorized parties. In that architecture, the data was exposed to the middleware as a trusted party, whereas in this paper we remove that trust by instead using a Trusted Execution Environment with remote attestation functionality.

\begin{figure}
  \centering
  \includegraphics[scale=.75]{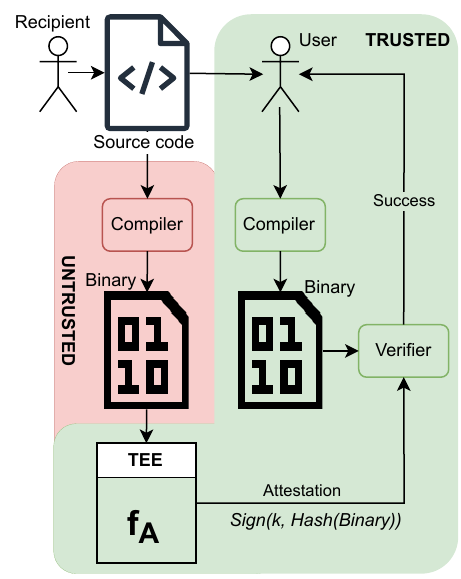}
  \caption{Overview of the attestation process. The recipient of the computation compiles and deploys the code on the TEE.
The source code is also made available to the users. Each user can inspect and
compile the code and by remote attestation verify that the compiled code running on the TEE matches the inspected source code.}
  \label{fig:attestation}
\end{figure}

\subsection{Trusted Execution Environments}
A trusted execution environment, or TEE, is a tamper-resistant processing environment that guarantees the integrity and confidentiality of its run-time state~\cite{teedef}. In addition, it provides a remote attestation procedure to prove trustworthiness for third parties. Common examples of such technologies are Intel SGX\footnote{https://www.intel.com/content/www/us/en/developer/tools/software-guard-extensions/overview.html} and AMD SEV\footnote{https://developer.amd.com/sev/}. This technology makes it possible to securely execute computations on a non-trusted computer. With remote attestation, a remote user also has the ability to ensure that one communicates with an actual TEE running a specific version of the code, which also gives integrity guarantees.

The concept of trusted execution platforms provides the possibility to keep many of the security guarantees of homomorphic encryption while keeping the computational and storage overhead low.
The core aspect of our system is that we transfer the transparency of the computation in the TEE directly towards every user, by allowing them to use remote attestation before transmitting any sensitive information.

Remote attestation is a process where the TEE gives a cryptographic proof of both the hardware platform itself and the software that is running~\cite{teeAttestation,remote-attestation-sgx-example,menetrey2022exploratory}. This proof consists of a cryptographically signed message that contains the hash of the currently running software, as well as information regarding the current hardware stack. The users then have the possibility to compare the hash of the binary with their own version by compiling the source code themselves. This requires that the build is reproducible~\cite{repBuilds}. For an overview of the attestation process, see Figure \ref{fig:attestation}.

\section{Architecture}
\label{sec:architecture}

The architecture of our solution consists of four parts; users that contribute data, a middleware that stores the data, a recipient that requires aggregated user data, and a TEE that performs the aggregation.
\new{In this section, we will cover all parts in detail, including the execution protocol.}

\begin{figure}
  \centering
  \includegraphics[width=0.5\textwidth]{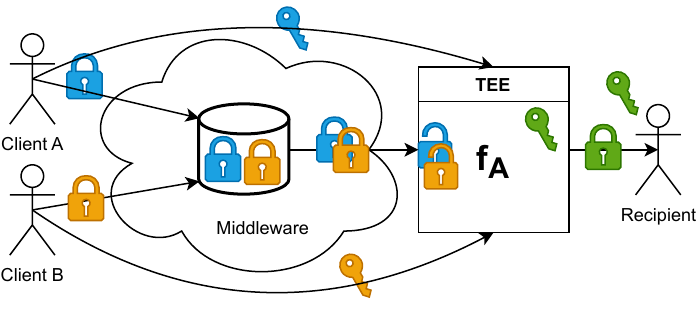}
  \caption{Overview of the confidential computing architecture. The architecture shows the flow of how data moves between users and the middleware, between the middleware and the TEE, and finally to the recipient. Note that decryption keys provided by the users to the TEE are sent using a separate channel directly between the user and the TEE.}
  \label{fig:arch-overview}
\end{figure}

\subsection{Overview}
Fig.~\ref{fig:arch-overview} shows a high-level overview of the architecture, which consists of the following elements:

\paragraph{Users}
Own the data and are the party that provides it to the cloud. They are responsible for generating their own encryption keys and providing only encrypted data to the cloud.
They are responsible for inspecting the code of the TEE (or delegating this task to a trusted entity) and, after remote attestation, for providing the TEE with the decryption key.

\paragraph{Middleware}
Is responsible for storing and releasing data.
It is also responsible for providing data to the TEE when requested to do so. 
In a real production setup, the middleware would be the party that starts the TEE on a per-request basis, but in our example, the TEE is assumed to be always running.

\paragraph{Trusted execution environment}
The TEE needs to provide security guarantees in terms of isolation and remote attestation, where the latter assumes that the code running on the TEE is publicly available.
The recipient that would like to receive aggregated data creates a function that is compiled and signed, as well as published for all participants to review. We call this function the \emph{aggregation function}, or $f_A$ for short. $f_A$ needs to have the following properties:
\begin{itemize}
  \item A public encryption key that is used to encrypt the result. This key needs to be tied to some entity so that participants can verify who can access the resulting data. This entity can, for example, be a known data aggregator that the participants can verify as an acceptable data handler.

  \item An API used for providing a decryption key tied to a specific user. We assume that all users are verified and unique.
  There should be no possibility for $f_A$ to leak any secret key.
  
\item Functionality to retrieve encrypted data and trigger the computation.
  
\item Functionality to return the result of the computation.
  
\end{itemize}


We use a hard-coded public key to encrypt the result so that the recipient of the aggregation function is known and fixed a priori, as recommended by Intel~\cite{remote-attestation-sgx-example}.
This allows any user to verify that only the party that possesses the corresponding private decryption key can access the result.
The users are the ones in possession of the keys and are the only ones who can provide them to the TEE. Hence, if a user does not explicitly permit to use their data, it cannot be used by the aggregation function.
As the enclave works with decrypted data, it can implement any functionality $f_A$.

\paragraph{Recipient}
The recipient is any party that requests aggregated data from data owners in the middleware. The recipient is responsible for setting up the TEE as well as publishing the code running on it, including their own public key. This makes sure that it is possible for any data owner to inspect the code and verify its correctness before giving it permission to use its private data.

\subsection{Execution protocol}
\begin{figure}
  \centering
  \includegraphics[width=0.5\textwidth]{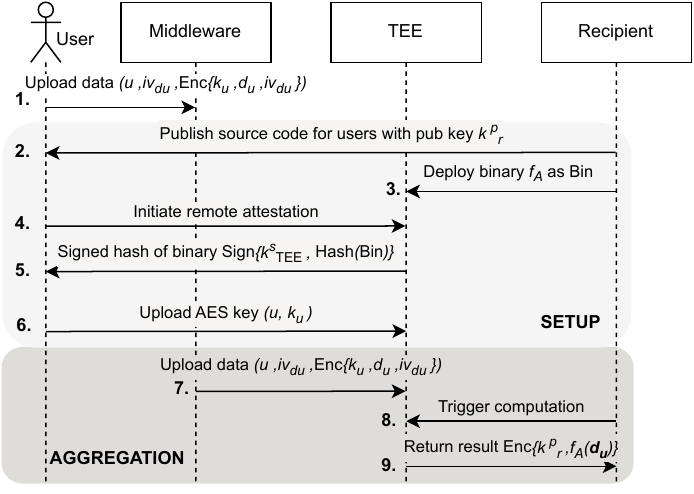}
  \caption{Sequence diagram showing the execution protocol.}
  \label{fig:seq-diagram}
\end{figure}

In Fig.~\ref{fig:seq-diagram}, we illustrate the steps involved in confidentially computing an arbitrary computation with consenting users' private data and sending the result to the recipient. The result of the computation is only available to the recipient and is not readable by either the users or the middleware.

We assume that all communication occurs over an authenticated, encrypted channel. (In our implementation, we use TLS to secure these channels.) To simplify the following explanation, we do not show the keys and encryption used to secure the communication channels and instead focus on the (additional) keys used to protect the data.
Without loss of generality, the following discussion covers an example where a user $u$ submits one data item $d_u$. In our system, it is also possible to submit data several times.

\paragraph*{1. Data upload}
We assume that every user $u$ already possesses a secret random AES key $k_u$ used for encrypting and decrypting their data. The users provide their data $d_u$ in encrypted format $\mathrm{Enc}(k_u, d_u, iv_{d_u})$, together with their identifier and the initialization vector $iv_{d_u}$ used to encrypt the data, to the middleware for backup and availability.

Note that for simplicity, in Figure \ref{fig:seq-diagram}, we denote the users' data $d_u$ as one data point. In general, this is not limited to any specific data type and could be anything from an integer, a float, an arbitrary string, or a tuple.
This step may occur several times and a user may upload new data when it becomes available, even though we show only one data upload here.

\paragraph*{2. Publish source code to users}
The recipient publishes the source code along with the public key $k^p_r$.
When the recipient has deployed the computation function on the TEE, the code is distributed among users.

The recipient can delegate code development to a third party.
This code can hence be inspected by each user. When inspecting the code, one should check that it does not leak confidential information such as private keys and data.
One may note that the code running in the enclave in general consists of two parts: a public interface for input and output of data, and the aggregation function itself. The first part should be standardized and should hence not need to be changed for different implementations of the aggregation function. The aggregation function should in general be short in terms of lines of code. It only needs to do as much for the aggregation to be considered non-confidential. Hence, it should be feasible to manually inspect the code in a rather quick and thorough manner. This also means that it should not need to be updated frequently. 

\paragraph*{3. Deploy binary}
Before any user can interact with the TEE, the recipient has to deploy the aggregation function $f_A$ on the TEE. The receiver hence compiles the same previously published source code to a binary, $\mathit{Bin}$, and uses this to set up the TEE.

\paragraph*{4. User initiates remote attestation}
When the user has determined that they want to participate, they initiate the remote attestation procedure by requesting proof from the TEE.

\paragraph*{5. Signed hash of binary}
The attestation consists of a signed hash $\mathrm{Sign}(k^s_\mathit{TEE}, \mathrm{Hash}(\mathit{Bin}))$ of the binary that is returned by the TEE. The hash is signed using a key derived by the TEE, $k^s_\mathit{TEE}$, and can be verified with the manufacturer of the TEE. The hash itself can be compared to the hash of the code that the users compile themselves. Hence, the user can be certain that the correct version of the code is running.



\paragraph*{6. User uploads AES key}
After a successful attestation procedure, the users can securely upload their key $k_u$ together with their identifier $u$ as a means of authorizing the TEE to include their data in the computation.
Since the user has inspected the code and is certain that the same code is running in the TEE, there should be no risk of any malicious activity.
Note that this step is done once and not necessarily by the same unit that provides data to the middleware, but by any device controlled by the user. A constrained device does not need to be able to attest the enclave, but only to be able to encrypt data and send it to the middleware.


\paragraph*{7. Middleware uploads encrypted data}
When enough users have authorized the enclave to use their data, the middleware can upload all encrypted data $\mathrm{Enc}(k_u, d_u, iv_{d_u})$, together with the initialization vectors $iv_{d_u}$ and the user ID's $u$, in one batch to the enclave. The enclave uses the previously provided keys to decrypt the data and store it in memory, isolated from the host.

\paragraph*{8. Trigger computation}
When the data is uploaded, the recipient can trigger the computation. Since the enclave now works with decrypted data, there are in general no limitations on what kind of computation can be created in the enclave, and the computational overhead is only limited to the enclave itself.

\paragraph*{9. Return encrypted result}

When the computation is done, the enclave uses the public key $k^p_r$ of the recipient to encrypt the aggregated result $f_A(\bold{d_u})$ before it is returned. This guarantees that only the entity in possession of the corresponding private decryption key can access the result.
Note here that $\bold{d_u}$ (bold) is denoting an array of multiple users' data, since the aggregation in general works with many different users' confidential data.




\section{\new{Threat analysis}}
\label{sec:security}
We define the following requirements for the system: R1) \textit{Data traceability:} It should be possible to verify who can access what data and which computations the data can be a part of; R2) \textit{Data aggregation:} It should be possible to compute arbitrary data aggregations over multiple mutually distrusting users without need to trust the service provider; R3) \textit{Flexibility:} It should be possible to dynamically update the participating users without changing the encryption; R4) \textit{Data confidentiality:} Confidential data should not be leaked to unauthorized parties.

\subsection{\new{Threat model}}
\new{
We assume that the middleware is not malicious, but could be compromised by a malicious party, enabling them  to observe and manipulate network traffic and data on the host computer. 

We assume that the TEE is running on an untrusted host, but that the hardware isolation is perfect. Mechanisms that try to bypass the isolation, e.\,g., via side-channel attacks~\cite{teeHardToTrust} or physical attacks, are outside the scope of this work. We also assume that the firmware on the TEE has not been tampered with.

The recipient can do anything with the data, hence the aggregation must not leak confidential data belonging to individual users. We further assume that the middleware and the recipient are independent and do not cooperate. If they did, it would be hard to protect against repeated queries of the encrypted data. Such attacks could be protected against by demanding that the aggregation is differentially private, which is left as future work.

Users are assumed to be untrusted in the sense that they may try to extract any data belonging to other users, but will not submit manipulated data to the middleware.
Detecting false data is orthogonal to our work and could, for example, be solved by anomaly detection~\cite{anomalyDetectionSurvey,anomaliesTimeseriesReview}.
}

\subsection{Security properties}
Since all data are stored in encrypted form in the middleware, the middleware cannot leak any data even if compromised. At the same time, since the users are in possession of their own decryption key, they can access their own data.
By the use of TEEs and aggregation functions, the user is in control of the data and can share and compute over arbitrary implementations of $f_A$ without the risk of leaking data to any unauthorized parties. By the means of remote attestation~\cite{remote-attestation-sgx-example,remote-attestation-sgx} the users can be sure that they provide their keys to a trusted environment running a specific version of $f_A$, which also gives the guarantees that the correct computation is carried out \new{as well as that the result is anonymous}.

The result of $f_A$ is protected by encrypting it with a hard-coded public key. Since it is hard-coded, the result-sharing function can be verified by remote attestation, and any user sending data to the enclave can be sure that only the party that possesses the corresponding decryption key can access the result.

\subsection{\new{Fulfillment of requirements}}
R1 is met by encrypting data with the user's own key before storing it in the cloud. 
Since the data only can be used if a user has provided the private key, one can verify which computations the data can be a part of.

R2 is met by computing over decrypted data. Since the data is decrypted before the aggregation, there are no limitations to what kind of computation can be implemented. Because of remote attestation, any user can inspect and verify the code before allowing their data to be used, hence it is possible to compute arbitrary computations on mutually distrusting users' data. At the same time, since the middleware only holds encrypted data, users do not need to trust it.

R3 is also met by being able to decrypt data in the enclave. Because of this, there is no need for secret-sharing among users before carrying out a computation. Any user can be a part of it at any time, using data collected at any time. This makes the system more flexible since no communication has to be done between users. In addition, any user can use the middleware as a secure means of storing data, and at a later stage decide to make the data available in the aggregation, without the need for changing any encryption or keys.

R4 is met using a combination of storing encrypted data and using remote attestation. Since data is stored in encrypted form, the middleware cannot leak any data. At the same time, remote attestation gives the user the possibility to inspect the code thoroughly before providing their key. Without the key, the data cannot be used.

\section{Implementation}
\label{sec:implementation}
To demonstrate that the system can be used in a practical setting, we have implemented a prototype illustrating its main aspects.\footnote{\new{\url{https://github.com/marbirg/sharing-without-showing}}}
The system uses Microsoft Azure's confidential computing environment with an SGX-enabled virtual machine (see Table \ref{table:vm-properties}). The architecture consists of three parts: the aggregation function, the middleware, and the users. The secure aggregation function consists of an untrusted host and a secure enclave.
We denote the machine running the aggregation function as the \emph{server}.


The server is written in C/C++ using the \emph{Open Enclave SDK}~\cite{openEnclaveSDK}. The Open Enclave SDK  is an open-source software development framework that
aims to generalize the development of enclave applications across TEEs from different hardware vendors.\footnote{\url{https://github.com/openenclave/openenclave}}

The application has been developed such that its build is reproducible.
Our implementation uses the AES algorithm in CBC mode with block size 16 and 256-bit random keys.

\begin{table}
  \caption{System properties for SGX-enabled VM running on Microsoft Azure Cloud.}
 \label{table:vm-properties}
  \centering
  \begin{tabular}{l p{55mm}}
\hline
    \textbf{Property} & \textbf{Value}\\ \hline
    OS/Kernel & Linux sgx 5.4.0-1104-azure \#110$\sim$18.04.1-Ubuntu SMP x86\_64 GNU/Linux \\
    CPU & Intel(R) Xeon(R) E-2288G CPU @ 3.70GHz \\
    Size & Standard DC1s v2 \\
    vCPUs & 1 \\
    RAM & 4\,GiB\\
\hline
  \end{tabular}

\end{table}

\subsection{TEE-server}
The TEE-server consists of a host and a secure enclave. The host is the layer that is running outside the TEE and invokes the trusted code running inside the enclave. This part is based on the \emph{Attested TLS sample}\footnote{\url{https://github.com/openenclave/openenclave/tree/master/samples/attested_tls}} provided by Open Enclave.
The enclave creates a web server with a secure \new{attested }TLS connection exposing an API for secure communication, including the functionality for remote attestation. \new{Hence, all parties that communicate with the TEE are using an attested TLS channel. }The enclave has one endpoint for retrieving a decryption key, one endpoint for retrieving encrypted data, and one endpoint for triggering the computation. In the last step, the result from the computation is returned in the response, encrypted with the recipient's public key. The encryption library used is \textit{Mbed TLS}.\footnote{\url{https://github.com/Mbed-TLS/mbedtls}}


The recipient's public encryption key is hard-coded (as recommended by Intel~\cite{remote-attestation-sgx-example}), and hence any change to it would be noticed when attesting the enclave.

\subsubsection{Remote attestation}
This functionality is provided by Intel SGX, and the procedure for interfacing with it is included in the Open Enclave framework. The procedure is carried out as part of the TLS handshake, which means that every connection to the enclave is attested before any data is sent.

\subsubsection{Key upload}
The keys are base64-encoded and sent to the server as JSON objects (on an attested TLS channel), together with the ID of the user. The server stores them as another JSON object that maps the IDs to the corresponding key.

\subsubsection{Data upload}
Each data point consists of the encrypted value (base64-encoded), together with the initialization vector used in the AES encryption process. The middleware can hence send all data in one batch to the server to make it available to be used in the aggregation. The server matches the ID sent with each data value with the corresponding key already provided and tries to decrypt the data. If it succeeds, this means that the user has authorized the enclave to use the data by providing their key. This data is then stored in the enclave's secure memory (which is isolated from the host).

\subsubsection{Aggregation}
The aggregation function is triggered using a separate endpoint.

Since the data stored in the enclave is now decrypted, the computation is done on plain text, and hence there are no limitations (except for available memory) on what kind of computation can be implemented.

For benchmarking, we implemented four different aggregation functions: a sum, least square fit, histogram, and SVM classification.

The result of the computation is encrypted with the hard-coded public key and returned in the response. This guarantees that only the entity in possession of the corresponding decryption key will be able to read the result from the computation.

\subsection{Client}
The users' clients generate their own AES keys that are used to encrypt the data before sending it to the middleware. Each data point is sent to the middleware as a tuple containing the encrypted data together with the initialization vector. All communication is done over a secure TLS channel.
In addition, each user also sends their decryption key to the TEE. When doing this, the client verifies that the TEE instance is the expected one by the remote attestation procedure, and then deploys the key using \new{the attested} TLS channel. The attestation procedure is done during the TLS handshake and is hence verified on every request.



\subsection{Middleware}
The middleware stores encrypted data from users so that it is backed up and available. It never receives any decryption keys and can hence never access the original data.
Upon request, the middleware can send this data in batches to the TEE for secure aggregation.

All communication between the user and the middleware uses TLS, \new{and all communication} between the middleware and the TEE \new{uses attested TLS}.



\subsection{Recipient}
The recipient is the party that wants the result of a pre-defined aggregation function. The recipient is responsible for creating this function, as well as making the source code available for inspection by any party that could be relevant for participating with their data. The recipient is also responsible for making this build reproducible.
When the middleware has uploaded the data to the enclave, the recipient can trigger the aggregation process and retrieve the encrypted result. Since the recipient is the party holding the private decryption key for the result, this is the only party that has the ability to access the result of the computation. \new{The communication between the recipient and the TEE is done on an attested TLS channel, hence the recipient can be sure that the correct computation has been carried out.}

\section{Evaluation}
\label{sec:evaluation}
We evaluate the performance of the system by implementing a variety of use cases and evaluating them in setups with and without the enclave.
%

For the experiments, we have used $5000$ data points pertaining to the confidential data of individual users. We have encrypted every data point with a $256$ bit random AES key before sending it to the enclave.


Our evaluation targets four different use cases. The first use case is a simple function that computes the sum of values that it receives as input. The second use case leverages the LIBSVM\footnote{\url{https://www.csie.ntu.edu.tw/~cjlin/libsvm/}} \footnote{\url{https://github.com/cjlin1/libsvm}} library to perform SVM (Support Vector Machine) classification. The third and the fourth use cases are based on the GNU Scientific Library\footnote{\url{https://www.gnu.org/software/gsl/}} (GSL). Here, we evaluate the functionality for computing a histogram over encrypted data, as well as for computing the least square fit on another encrypted dataset.

All our use cases are implemented using the Open Enclave framework, and for all use cases, there is a plain service running equivalent code but performing the same computation outside an enclave and working with unencrypted data. Both the secure enclave version and the non-secure plain version communicate on a TLS channel.



\subsection{Core functions}
We have implemented four different functions for evaluation; a sum, an SVM classification, a histogram classification, and a least square fit. In all cases, the data is uploaded in one step, where the enclave performs decryption and saves the data for later computation. In the trigger stage, the actual computation is triggered. Here, the data is loaded from memory, formatted based on the triggered function, and the result is returned in the response. For the service running in the enclave, the result is always encrypted using the recipient's public key before being returned. In the plain service, the result is returned in plain text. 
In the following, we provide a detailed description of these functions.

\subsubsection{Confidential sum computation}
This function takes an array of integers and returns the sum. It is implemented in C++ without external libraries. 
In both the plain and the secure versions, previously uploaded data values are loaded from memory, converted to integers, and then added. In the plain service, the result is returned as is, while in the secure enclave version, it is first encrypted using the recipient's public key.


\subsubsection{Secure SVM classification}
To show that the system can work with existing off-the-shelf software in more complex settings, we evaluate it by running a version of LIBSVM inside the enclave. LIBSVM is a framework for SVM training and classification and has been used in other works for benchmarking performance in  TEEs~\cite{benchmarkSGX}.
By using the LIBSVM framework, we implement a proof of concept for classifying breast cancer data.




\paragraph{Training}
We have trained the model outside the enclave using the \emph{breast-cancer}\footnote{\url{https://www.csie.ntu.edu.tw/~cjlin/libsvmtools/datasets/binary.html\#breast-cancer}\label{fn:breastCancer}} dataset provided on the LIBSVM web page. Then, we compiled the trained model into the library and used it for classification tasks.
Note that the main reason for training the model outside the enclave is for transparency toward the users that provide data. If the model were loaded into the enclave after initiation, or trained inside the enclave, it would not be possible for any user to know how the classification would be carried out by means of remote attestation. A malicious middleware or recipient could train the model in such a way that it revealed users' confidential raw data. 

\paragraph{Classification}
Data points not used for training were uploaded to the enclave for classification. The service uses the trained model to classify each data point for each user. The result of all classifications is then returned.


\subsubsection{Histogram}
To further show the generality of the system, we incorporate the GNU Scientific Library (GSL). This library is implemented in C/C++ and is in active development. It consists of many tools for general scientific computations, including a large library for statistical computations.


To compute the histogram, we use \emph{gsl\_histogram\_alloc} to initiate a new histogram object, the \emph{gsl\_histogram\_set\_ranges\_uniform} to initiate the bins, and iteratively call \emph{gsl\_histogram\_increment} for each value. The result is returned as a comma-separated list.

\subsubsection{Least square fit}
The LSF function can be used to interpolate a linear function from different data points, for example,  to find correlations between data.

The input data is encoded as pairwise comma-separated 2D coordinates. These are extracted and encoded as $x$ and $y$ arrays. By using the GSL-provided function \texttt{gsl\_fit\_linear}, we compute the coefficients, which are returned as a comma-separated list. 

This kind of application can be used to find correlations in confidential data without actually being able to access any raw data from any individual user.



\subsection{Data generation}
For all functions but the SVM classification, we have generated the data randomly and have encrypted them with individual keys.

For the SVM use case, the dataset consists of $683$ data points, where $600$ points were randomly selected and used for training and the remaining $83$ points were used for testing and evaluation. Note that this means that the same data point has been sampled multiple times in the experiments, without loss of generality.




\subsection{Benchmark}


Both the client and the server are running on the same virtual machine in the Microsoft Azure cloud environment. The same virtual machine acts as a host for both the secure enclave and the non-secure plain service.

We have developed a benchmark framework in Python
to facilitate the quick generation of different kinds of data, as well as to measure the computation time.
This code can generate data suitable for all our use cases, create and provide encryption keys for the mocked users, as well as encrypt data before sending it to the enclave.



The benchmark consists of comparing the time for uploading previously deployed data between the secure and the plain service, as well as the actual computation time for the two variants.
The time is measured between starting the request and receiving the response.
Each round consists of sending a batch of users' values from the middleware to the aggregation service and recording the time until the service responds that the data has been stored successfully. We measure the time between triggering the aggregation and the result being returned.

\subsection{Time comparison}
The different computation times for the plain service and the enclave are mainly dependent on three aspects: the overhead generated by the enclave itself,  the time spent by the enclave to decrypt the incoming data and encrypt the output, and the amount of data to send to the service. In Table \ref{tab:enc-plain-result}, we present the measured time to run the different applications. The table shows the time for uploading data, the time for computing the actual function, the sum of these processes as well as the standard deviation (taken on the total time). We also show the geometric mean of the two services and an overhead of $1.62$.
When comparing the data upload time for the different cases we can see that we have an overhead of $1.43$ for the sum, $1.76$ for the SVM, $1.78$ for the LSF, and $1.51$ for the histogram. The corresponding overhead for the aggregation is $1.84$ for the sum, $1.56$ for the SVM, $2.07$ for the LSF, and $1.84$ for the histogram.

When taking into account the increased security and confidentiality guarantees that this system gives, we consider this overhead acceptable. In fact, existing cloud-based IoT platforms such as IFTTT are allowed up to 15 minutes
to execute\footnote{https://ifttt.com/docs/applets}, which indicates that our solution is applicable to real-world settings.

\subsection{Memory comparison}
The difference in memory consumption is mainly dependent on two aspects; 1) different amounts of data one needs to send to the aggregation service depending on whether one uses the secure enclave or the unencrypted service, and 2) different amounts of data that the service needs to store.

In the first case, one needs to send a larger amount of data to the secure aggregation service, since encrypted data in general takes up more space than unencrypted data.
In our case, we have used keys of size $256$ bits, and hence our cipher text will be at least $32$ bytes. In the benchmark, we are only using small numbers (a few bytes) to send to the service, so for the unencrypted case, we will send the actual bytes per value. For the secure encrypted case, each value will be sent as a $32$-byte string. Note that this has a large effect when working with small values, while the difference can be close to zero for sending values that are in the size of $32$ bytes.

In the second case, for the secure service, we also need to provide the AES key, which the service needs to store. The key transportation is not really an issue, since it is only done once, in the setup phase. Yet the service needs to be able to store the private keys from all users as well as an identifier connected to each key, something that is not necessary in the plain unencrypted case. 

\begin{table}
    \caption{Evaluation of computation time [ms] for different applications using encrypted data inside an enclave and the plain equivalent service. Evaluated over $100$ iterations using data from $5000$ simulated users.}
    \label{tab:enc-plain-result}
\begin{tabular}{llrrrr}
\hline
 \textbf{Function}   & \textbf{Type}   &   \textbf{Upload} &  \textbf{Computation} &   \textbf{Total} &   $\sigma$ \\
&&\textbf{mean}&\textbf{mean}&&
\\
\hline
 Sum        & Enclave &         1519 &            59 &    1578 &    46 \\
            & Plain   &         1050 &            32 &    1082 &    20 \\
 SVM        & Enclave &         2142 &           126 &    2268 &    27 \\
            & Plain   &         1216 &            80 &    1296 &    25 \\
 LSF        & Enclave &         1722 &            56 &    1779 &    29 \\
            & Plain   &          970 &            27 &     997 &    16 \\
 Histogram  & Enclave &         1509 &            46 &    1555 &    36 \\
            & Plain   &          998 &            25 &    1023 &    19 \\
\hline    
 \multicolumn{6}{c}{\textbf{Geometric mean: }
Enclave: $1773$; Plain: $1094$; Overhead: $1.62$}\\

\end{tabular}
\end{table}
\def\maxOverhead{1.78}

\subsection{Taxi distribution use case}
To show how simple functions as the above could be put to use in a practical full-scale application, we have developed a setup to confidentially compute the distribution of taxis in a medium-sized city. In this use case, we simulate the positions of taxis from different companies. Each taxi has an encryption key as well as the capability to detect its own position. In addition, they can encrypt this position and send it to the middleware in real-time. 

For this use case, we measure the time it takes to upload the data to the enclave as well as the time to retrieve the computed distribution (see Table~\ref{tab:dist-comp-time}).
For the experiment, we set up the client and the TEE on different machines, and communication was carried out over the public Internet. 
The table shows the time it takes to upload the encrypted data to the TEE, the time it takes to trigger the computation and retrieve the result, as well as the total time and the standard deviation. As one can see, the computation time is barely affected by the number of taxis, while the upload time grows linearly with the size of the data set.

In a simulation with $5000$ taxis, it takes about $14\,\mathrm{s}$ to send all encrypted data to the TEE and retrieve the distribution. Since this computed result is assumed to be public, the middleware can buffer it and release the same computation to multiple individuals from the public and hence handle the scaling on that level.


In the implementation, we have modeled the city as a quadratic grid of size $1000\times1000$. At each tenth coordinate, there is an intersection where the cars randomly change direction. We accordingly split the map into $100\times100$ zones and the aggregation function counts the number of taxis in each zone.

This computation makes it possible for anyone to view the current distribution of taxis in the city. Information that can be used to detect traffic congestion or where it is easiest to get a taxi right now.
The companies can see the real-time position of all their cars but no other taxi company's cars. They could further use the distribution to know where it is best to send their cars. This is without having to trust the middleware with the sensitive position data unencrypted.




\begin{table}
    \caption{Evaluation of computation time [s] for the distribution of taxis in a city. Evaluated over $100$ iterations.}
\label{tab:dist-comp-time}
    \centering
\begin{tabular}{rrrrr}
  \hline
  \textbf{Taxis} & \textbf{Upload mean} & \textbf{Computation mean} & \textbf{Total} & $\sigma$\\ \hline
  100  & 0.35  & 0.35 & 0.70  & 0.17\\
  1000 & 2.79  & 0.34 & 3.14  & 0.39\\
  2000 & 5.65  & 0.37 & 6.01  & 0.55\\
  3000 & 8.53  & 0.38 & 8.92  & 0.64\\
  4000 & 11.01 & 0.39 & 11.40 & 0.79\\
  5000 & 13.78 & 0.41 & 14.19 & 0.69\\
  \hline
\end{tabular}
\end{table}


\subsection{Threats to validity}
In the current system, the enclave is assumed to always be running. Since a TEE has an increased cost compared to a regular server, this might not be the case in a real-world setting. This would affect the initial startup time for the computation, but foremost how keys are handled in the enclave. To be able to provide keys only once, these keys need to be stored in persistent storage, so-called 'seals'.\footnote{https://github.com/openenclave/openenclave/blob/master/samples/data-sealing/README.md} This has not been evaluated and could affect the performance of the secure service if keys are loaded from encrypted storage on disk rather than from the enclaves isolated memory.

\subsection{Research questions revisited}
\paragraph{\textbf{RQ1:} \rqa}
\newp{All data stored in the Middleware are encrypted with users' individual keys. Only an enclave with the right key can access that specific user's data.}
In addition, since the enclave is computing over decrypted data, \new{any computation can be implemented}.
\new{Because} of remote attestation, the user can be certain that the promised function will be carried out. At any point in time, a user can agree to use their data in a specified enclave without any requirement on how that data was encrypted or when it was collected. Hence, all our requirements are fulfilled.

\paragraph{\textbf{RQ2:} \rqb}
Remote attestation guarantees that the correct code is executed on the data. Hence, every user has the ability to verify that the function is not malicious. All data are encrypted and cannot be read \new{without} the decryption keys. As long as every party follows the protocol, the recipient accessing the aggregated data can be sure that the correct function has been computed. In the case when any party deviates from the protocol, for example, if users provide false data or the middleware sends the wrong data, this might give faulty aggregation results but does not leak any confidential data. 

\paragraph{\textbf{RQ3:} \rqc}
In our experiments, we have found an overhead of about $1.62$ when comparing the geometric mean over all use cases between the two variants. Since the secure system provides a lot more security and confidentiality guarantees for both the owners of the confidential data and the recipient that requests the resulting aggregation, we believe this to be acceptable.


\subsection{Discussion}
The experimental evaluation of our prototype implementation exhibits a worst-case overhead of about $\maxOverhead\times$ for the LSF fit computation (incl. data upload time).
Similar works that use TEEs~\cite{walnut, MlTrainingTee, privPresMlTee} compare their runtime with an equivalent system running outside a TEE, while we compare it to a setup with neither TEE nor encryption. The reason for this is that encryption would not be relevant without the TEE. The work of Islam et al.~\cite{secureIotAnalysis} follows a similar setup to achieve an overhead of $3.3\times$.  
Our rationale for comparing with a plain text model running outside of an enclave is driven by the use case of cloud-based IoT applications~\cite{DBLP:journals/ieeesp/BalliuBS19}. Currently,  most analytics on IoT data follow this setup, thus justifying the relevance of our comparison. Moreover, our system enables users to upload data at any time, while this data can be used at a later time in any kind of analytics.

When only looking at the overhead from the TEE, an overhead of about $1.2\times$ has been reported in the literature~\cite{performanceOfIntelSgx}. The rest of our overhead therefore most probably comes from decrypting and encrypting data (including extra transmission time related to sending more bytes). Because of the extra layer of both security and confidentiality this provides, we believe it is a very reasonable overhead.

Our focus has been on protecting the confidentiality of users' data, while still providing the possibility of performing computations on data from different mutually-distrusting users. Our \new{threat} model does not assume that the middleware is malicious in its construction, but that it might be compromised by an external party. Hence, we do not want to store plain text data on the middleware but can rely on it to be honest and follow the protocol.

We have implemented the system on Intel SGX, but any TEE could be used for our architecture.


\section{Related work}
\label{sec:relatedwork}

\def\secIot{Secure IoT Data Analytics}
\def\secML{Privacy-preserving ML}
\def\yao2pc{with Yao's protocol}

\def\yes{{\color{Green} \checkmark}}
\def\no{{\color{red} $\times$ }}

\begin{table*}
  \caption{Comparison of related work}
  \label{tab:relatedWork}
{\small
  \begin{tabular}{lllllll}
\hline
    \textbf{Work}                     & \textbf{Offline}                       & \textbf{Arbitrary}    & \textbf{Data}         & \textbf{Integrity} & \textbf{Multiple}    & \textbf{Dynamically}\\
                                      & \textbf{computation}                   &\textbf{computation}               & \textbf{protection}   & \textbf{protection} & \textbf{users}       & \textbf{add users} 
\\\hline
    Our work                         & \yes                                       & \yes                           & \yes                  & \yes                & \yes                 & \yes            \\ 
    \secIot~\cite{secureIotAnalysis} & \yes                                       & \yes                           & \yes                  & \yes                & \no                  & \no             \\ 
    Walnut~\cite{walnut}             & \no                                        & \yes (\yao2pc)                 & \yes                  & \yes                & \yes (2 users)       & \no             \\
    eTap~\cite{eTap}                 & Partly    & \yes                           & \yes                  & \yes                & \no                  & \no             \\ 
    minTap~\cite{minTap}             & \yes                                       & \yes                           & Partly                & \no                 & \no                  & \no             \\ 
    \secML~\cite{mpcMachineLearning} & \no                                        & \no                            & \yes                  & \yes                & \yes                 & \no             \\ 
    BeeKeeper 2.0~\cite{beekeeper}   & \yes                                       & \yes                           & \yes                  & \yes                & \no                  &  \no            \\ 
\hline
 
  \end{tabular}
}
\end{table*}










This section discusses our contributions in the context of existing works that address the problem of secure data analytics in the cloud.

\subsection{TEE-backed analytics}
%





Trusted execution environments have been used in many practical settings and have been the focus of many security- and privacy-enhanced prototypes. In the area of IoT, past works have mainly been focused around a single-user use case~\cite{secureIotAnalysis,iotBlockChTEE,walnut,privPresIotSgx}\new{,~\cite{streambox}}, often in a Trigger-Action Platform setting. Other works in the area mainly focus on protecting the device itself~\cite{iioteed,teeCloudFog,3rdParTEE}, something that is orthogonal to our work. In a survey paper investigating common usages for Intel SGX in an IoT setting~\cite{intelSgxSystematicStudy}, few papers presented a system for multiple users, and none focused on the specific case of aggregating confidential data from multiple mutually distrusting users. \new{In terms of MPC, this is the focus of \cite{mpcSGX}, but requires the users to be online during computation or trust a third party with their data. The work of~\cite{iron} could be extended to work on multiple mutually distrusting users but would require key sharing among all users.}

Machine learning is another area where trusted execution platforms have been popular. In most of these cases, the TEE is mainly used for either protecting the machine learning model itself~\cite{MlTrainingTee,privPresMlTee,machineLearningTEE} or, when using federated machine learning, to protect the integrity of the computation~\cite{flateeFedLearningTEE,fedLearningSchemeTee}. In this context, our work differs by having as the main focus the confidentiality of the user's sensitive data, while the model is considered public.

\new{TEEs has also been used to protect distributed computations~\cite{vc3,ryoan,opaque}, still in a single-user setting.}



\subsection{Trigger-Action Platforms}
Trigger-action platforms~\cite{iftttVsZapier} are very popular tools for user automation tasks in the cloud but can lead to the leakage of confidential data or misuse of services~\cite{fixTap,DBLP:journals/ieeesp/BalliuBS19}. Works that have focused on these issues include  eTap~\cite{eTap} and dTap~\cite{dTap18}, which require proof before allowing an action, minTap~\cite{minTap} that minimizes the amount of data sent to the service while keeping all functionality, and Sandtrap~\cite{sandtrap} that creates a sandbox environment using JavaScript.
All these works focus on the single-user setting and require users to trust the cloud provider with their sensitive data.

\subsection{Machine learning}
Machine learning in general, and federated learning~\cite{surveyFedLearning} in particular, are commonly used tools in many applications that handle confidential data. In some cases, the data used to train or update a model can be confidential~\cite{fedLearningBreastCancer,fedLearningHealthcare}, which has been the focus of many works using confidential data to update a model. These models usually work with complex secret sharing schemes and require the clients to be online during computation~\cite{mpcMachineLearning, flateeFedLearningTEE,flChainedMpc}.

We refer to the survey by Mo et al.~\cite{machineLearningTEE} for an overview of machine learning for confidential computing.

\subsection{Homomorphic encryption}
Several works have used homomorphic encryption~\cite{homEncSpringer} as a means to securely compute over data on the cloud~\cite{fheSurvey}. BeeKeeper~\cite{beekeeper} \new{uses} a blockchain to evaluate data using homomorphic encryption. eTAP~\cite{eTap} enables current TAP systems to evaluate triggers on encrypted data. Neither of these services has the ability to handle multi-user settings. Bonawitz et al.~\cite{mpcMachineLearning} consider multi-user settings in the context of privacy-preserving machine learning. Their system requires the users to be online during the computation and uses a complex schema of secret sharing. 

\subsection{Summary}
In Table \ref{tab:relatedWork}, we compare our system with existing works trying to create similar functionality of a system that protects the users' confidential data from a central system that performs aggregations. With \emph{Offline computation}, we mean that the user does not need to be online during the computation. The outlier here is eTap~\cite{eTap}, where the user does not need to be online during the computation, but need to update the scrambled circuits before any new trigger is made. We also compare the capabilities for \emph{Arbitrary computations}. Here, the work by Bonawitz et al.~\cite{mpcMachineLearning} is the only work that does not have that capability. In regards to \emph{Data protection} and \emph{Integrity protection}, minTap~\cite{minTap} is the only one that does not fulfill this completely. \emph{Multiple users} represent work that
is suitable to aggregate over data belonging to multiple individuals. Here, our work is the only one that can handle an arbitrary number of users. Our work is also the only work where it is possible to add users without exchanging keys with already enrolled users. \new{We believe that this is an important aspect when collecting statistics over a longer time period in a multi-user setting. New users can easily be added and improve the statistics since existing users do not need to be active when this happens.} 

As one can see, all works focus on protecting the users' data, and in general succeed. Our work stands out in this comparison as the only one where we can protect both data and integrity during the computation, as well as performing arbitrary computations on multiple mutually distrusting users while no user needs to be online during the computation.
In other words, several works fulfill parts of our requirements, but none fulfills them all.




\section{Conclusion}
\label{sec:conclusion}
We have proposed a system for carrying out confidential multi-user computation without complex secret sharing schemas and instead using a TEE. Our implementation uses Open Enclave as a means of interfacing with Intel SGX and off-the-shelf software to implement various sets of different computational functionalities. We have compared our secure computation with an equivalent service running outside the enclave and computing over unencrypted data. This setup is designed to find the worst-case overhead.

For the selected applications, we have an overhead of about $1.6$ times compared with using plain data. 
The selected functions have been chosen to show the generality of the system in different contexts, as well as to show that it is possible to use available frameworks with minimal (if any) modifications.

\new{We have foremost used an IoT setting in this paper since it works well for low-powered devices that are not always connected to the Internet, while such devices also can collect data of confidential nature. With that said,} our system can be used in multiple settings where sensitive data need to be both stored in the cloud and potentially later used for analytics, without deciding a priori which data or which users want to be a part of the computation.

\subsection*{Future work}
In our system, we have assumed that the middleware and the recipient do not cooperate to extract individual data from the user. If they did, the requirements for the computational function would be higher, to prevent that.
It is also the case that in the current system, the user is responsible for verifying that the function running in the enclave cannot leak private keys and that the function itself aggregates data in such a way that there is no issue with releasing the result to the recipient. It would be beneficial to use tools to verify certain aspects automatically, such as tools based on information flow control, differential privacy, or statistics.
It would also be beneficial to model the protocol in a semi-automatic prover such as Tamarin~\cite{meier2013tamarin}.
Such tools could be used to verify the absence of data leakage or give some certainty as to how much information a certain computation can leak.


\section*{Acknowledgements}
This work was supported by the Centre for Cyber Defence and Information Security (CDIS).

\balance

\bibliography{references}{}

\begin{thebibliography}{10}

\bibitem{notConnectingSmartDev}
Appliance makers sad that 50\% of customers won’t connect smart appliances.
\newblock \url{
  https://arstechnica.com/gadgets/2023/01/half-of-smart-appliances-remain-disconnected-from-internet-makers-lament/
  }.
\newblock Accessed: 2023-01-26.

\bibitem{iotSecBreach4}
{IoT} security breaches: 4 real-world examples.
\newblock \url{
  https://conosco.com/industry-insights/blog/iot-security-breaches-4-real-world-examples
  }, 2021.
\newblock Accessed: 2023-03-23.

\bibitem{openEnclaveSDK}
{Open Enclave SDK}.
\newblock \url{https://openenclave.io/sdk/}, November 2022.

\bibitem{sandtrap}
Mohammad~M. Ahmadpanah, Daniel Hedin, Musard Balliu, Lars~Eric Olsson, and
  Andrei Sabelfeld.
\newblock {SandTrap}: Securing {JavaScript}-driven trigger-action platforms.
\newblock In {\em 30th {USENIX} Security Symposium ({USENIX} Security 21)},
  pages 2899--2916. {USENIX} Association, August 2021.

\bibitem{fedLearningHealthcare}
Rodolfo~Stoffel Antunes, Cristiano Andr\'{e}~da Costa, Arne K\"{u}derle,
  Imrana~Abdullahi Yari, and Bj\"{o}rn Eskofier.
\newblock Federated learning for healthcare: Systematic review and architecture
  proposal.
\newblock {\em ACM Trans. Intell. Syst. Technol.}, 13(4), may 2022.

\bibitem{iotBlockChTEE}
Gbadebo Ayoade, Vishal Karande, Latifur Khan, and Kevin Hamlen.
\newblock Decentralized {IoT} data management using blockchain and trusted
  execution environment.
\newblock In {\em 2018 IEEE International Conference on Information Reuse and
  Integration (IRI)}, pages 15--22, 2018.

\bibitem{mpcSGX}
Raad Bahmani, Manuel Barbosa, Ferdinand Brasser, Bernardo Portela, Ahmad-Reza
  Sadeghi, Guillaume Scerri, and Bogdan Warinschi.
\newblock {Secure Multiparty Computation from SGX}.
\newblock In Aggelos Kiayias, editor, {\em Financial Cryptography and Data
  Security}, pages 477--497, Cham, 2017. Springer International Publishing.

\bibitem{DBLP:journals/ieeesp/BalliuBS19}
Musard Balliu, Iulia Bastys, and Andrei Sabelfeld.
\newblock {Securing IoT} apps.
\newblock {\em {IEEE} Security {\&} Privacy Magazine}, 2019.

\bibitem{secAwareIot}
Marcus Birgersson, Cyrille Artho, and Musard Balliu.
\newblock Security-aware multi-user architecture for {IoT}.
\newblock In {\em 2021 IEEE 21st International Conference on Software Quality,
  Reliability and Security (QRS)}, pages 102--113, 2021.

\bibitem{anomaliesTimeseriesReview}
Ane Bl\'{a}zquez-Garc\'{\i}a, Angel Conde, Usue Mori, and Jose~A. Lozano.
\newblock A review on outlier/anomaly detection in time series data.
\newblock {\em ACM Comput. Surv.}, 54(3), apr 2021.

\bibitem{mpcMachineLearning}
Keith Bonawitz, Vladimir Ivanov, Ben Kreuter, Antonio Marcedone, H.~Brendan
  McMahan, Sarvar Patel, Daniel Ramage, Aaron Segal, and Karn Seth.
\newblock Practical secure aggregation for privacy-preserving machine learning.
\newblock In {\em Proceedings of the 2017 ACM SIGSAC Conference on Computer and
  Communications Security}, CCS '17, page 1175–1191, New York, NY, USA, 2017.
  Association for Computing Machinery.

\bibitem{anomalyDetectionSurvey}
Varun Chandola, Arindam Banerjee, and Vipin Kumar.
\newblock Anomaly detection: A survey.
\newblock {\em ACM Comput. Surv.}, 41(3), jul 2009.

\bibitem{fixTap}
Xuyang Chen, Xiaolu Zhang, Michael Elliot, Xiaoyin Wang, and Feng Wang.
\newblock Fix the leaking tap: A survey of trigger-action programming ({TAP})
  security issues, detection techniques and solutions.
\newblock {\em Computers Security}, 120:102812, 2022.

\bibitem{fedLearningSchemeTee}
Yu~Chen, Fang Luo, Tong Li, Tao Xiang, Zheli Liu, and Jin Li.
\newblock A training-integrity privacy-preserving federated learning scheme
  with trusted execution environment.
\newblock {\em Information Sciences}, 522:69–79, 2020.

\bibitem{minTap}
Yunang Chen, Mohannad Alhanahnah, Andrei Sabelfeld, Rahul Chatterjee, and
  Earlence Fernandes.
\newblock Practical data access minimization in {Trigger-Action} platforms.
\newblock In {\em 31st USENIX Security Symposium (USENIX Security 22)}, pages
  2929--2945, Boston, MA, August 2022. USENIX Association.

\bibitem{eTap}
Yunang Chen, Amrita~Roy Chowdhury, Ruizhe Wang, Andrei Sabelfeld, Rahul
  Chatterjee, and Earlence Fernandes.
\newblock Data privacy in trigger-action systems.
\newblock In {\em 2021 IEEE Symposium on Security and Privacy (SP)}, pages
  501--518, 2021.

\bibitem{remote-attestation-sgx-example}
Intel Corporation.
\newblock Code sample: Intel® software guard extensions remote attestation
  end-to-end example.
\newblock \url{
  https://www.intel.com/content/www/us/en/developer/articles/code-sample/software-guard-extensions-remote-attestation-end-to-end-example.html
  }.
\newblock Accessed: 2022-10-12.

\bibitem{remote-attestation-sgx}
Intel Corporation.
\newblock Intel® software guard extensions.
\newblock \url{
  https://www.intel.com/content/www/us/en/developer/tools/software-guard-extensions/attestation-services.html
  }.
\newblock Accessed: 2022-10-12.

\bibitem{DBLP:journals/iacr/CostanD16}
Victor Costan and Srinivas Devadas.
\newblock Intel {SGX} explained.
\newblock {\em {IACR} Cryptol. ePrint Arch.}, page~86, 2016.

\bibitem{fedLearningBreastCancer}
Saloni Dagli, Kashvi Dedhia, and Vinaya Sawant.
\newblock A proposed solution to build a breast cancer detection model on
  confidential patient data using federated learning.
\newblock In {\em 2021 IEEE Bombay Section Signature Conference (IBSSC)}, pages
  1--6, 2021.

\bibitem{dTap18}
Earlence Fernandes, Amir Rahmati, Jaeyeon Jung, and Atul Prakash.
\newblock Decentralized action integrity for trigger-action {IoT} platforms.
\newblock In {\em 22nd Network and Distributed Security Symposium (NDSS 2018)},
  February 2018.

\bibitem{iron}
Ben~A. Fisch, Dhinakaran Vinayagamurthy, Dan Boneh, and Sergey Gorbunov.
\newblock Iron: Functional encryption using {Intel SGX}.
\newblock Cryptology ePrint Archive, Paper 2016/1071, 2016.
\newblock \url{https://eprint.iacr.org/2016/1071}.

\bibitem{privPresIotSgx}
Pascal Gremaud, Arnaud Durand, and Jacques Pasquier.
\newblock {Privacy-Preserving} {IoT} cloud data processing using {SGX}.
\newblock In {\em Proceedings of the 9th International Conference on the
  Internet of Things}, IoT 2019, New York, NY, USA, 2019. Association for
  Computing Machinery.

\bibitem{ryoan}
Tyler Hunt, Zhiting Zhu, Yuanzhong Xu, Simon Peter, and Emmett Witchel.
\newblock {Ryoan: A Distributed Sandbox for Untrusted Computation on Secret
  Data}.
\newblock In {\em 12th USENIX Symposium on Operating Systems Design and
  Implementation (OSDI 16)}, pages 533--549, Savannah, GA, November 2016.
  USENIX Association.

\bibitem{secureIotAnalysis}
Md~Shihabul Islam, Mustafa~Safa {\"{O}}zdayi, Latifur Khan, and Murat
  Kantarcioglu.
\newblock Secure {IoT} data analytics in cloud via {Intel} {SGX}.
\newblock In {\em 13th {IEEE} International Conference on Cloud Computing,
  {CLOUD} 2020, Virtual Event, 18-24 October 2020}, pages 43--52, 2020.

\bibitem{3rdParTEE}
Jinsoo Jang and Brent~Byunghoon Kang.
\newblock 3rdpartee: Securing third-party iot services using the trusted
  execution environment.
\newblock {\em IEEE Internet of Things Journal}, 9(17):15814--15826, 2022.

\bibitem{pverifier}
Ze~Jin, Luyi Xing, Yiwei Fang, Yan Jia, Bin Yuan, and Qixu Liu.
\newblock P-verifier: Understanding and mitigating security risks in
  cloud-based {IoT} access policies.
\newblock In {\em Proceedings of the 2022 ACM SIGSAC Conference on Computer and
  Communications Security}, CCS '22, pages 1647--1661, New York, NY, USA, 2022.
  Association for Computing Machinery.

\bibitem{benchmarkSGX}
Sandeep Kumar, Abhisek Panda, and Smruti~R. Sarangi.
\newblock A comprehensive benchmark suite for {Intel SGX}.
\newblock arXiv, 2022.

\bibitem{repBuilds}
Chris Lamb and Stefano Zacchiroli.
\newblock Reproducible builds: Increasing the integrity of software supply
  chains.
\newblock {\em IEEE Software}, 39(2):62--70, 2022.

\bibitem{homEncSpringer}
Ninghui Li.
\newblock {\em Homomorphic Encryption}, pages 1320--1320.
\newblock Springer US, Boston, MA, 2009.

\bibitem{flChainedMpc}
Yong Li, Yipeng Zhou, Alireza Jolfaei, Dongjin Yu, Gaochao Xu, and Xi~Zheng.
\newblock Privacy-preserving federated learning framework based on chained
  secure multiparty computing.
\newblock {\em IEEE Internet of Things Journal}, 8(8):6178–6186, Apr 2021.

\bibitem{teeHardToTrust}
Weijie Liu, Hongbo Chen, XiaoFeng Wang, Zhi Li, Danfeng Zhang, Wenhao Wang, and
  Haixu Tang.
\newblock Understanding {TEE} containers, easy to use? {Hard} to trust, 2021.

\bibitem{fheSurvey}
Chiara Marcolla, Victor Sucasas, Marc Manzano, Riccardo Bassoli, Frank H.~P.
  Fitzek, and Najwa Aaraj.
\newblock Survey on fully homomorphic encryption, theory, and applications.
\newblock {\em Proceedings of the IEEE}, 110(10):1572--1609, 2022.

\bibitem{meier2013tamarin}
Simon Meier, Benedikt Schmidt, Cas Cremers, and David Basin.
\newblock The {TAMARIN} prover for the symbolic analysis of security protocols.
\newblock In {\em Computer Aided Verification: 25th International Conference,
  CAV 2013, Saint Petersburg, Russia, July 13-19, 2013. Proceedings 25}, pages
  696--701. Springer, 2013.

\bibitem{menetrey2022exploratory}
J{\"a}mes M{\'e}n{\'e}trey, Christian G{\"o}ttel, Marcelo Pasin, Pascal Felber,
  and Valerio Schiavoni.
\newblock An exploratory study of attestation mechanisms for trusted execution
  environments.
\newblock {\em arXiv preprint arXiv:2204.06790}, 2022.

\bibitem{teeAttestation}
Jämes M{\'{e}}n{\'{e}}trey, Christian Göttel, Anum Khurshid, Marcelo Pasin,
  Pascal Felber, Valerio Schiavoni, and Shahid Raza.
\newblock Attestation mechanisms for~trusted execution environments
  demystified.
\newblock In {\em Distributed Applications and Interoperable Systems}, pages
  95--113. Springer International Publishing, 2022.

\bibitem{machineLearningTEE}
Fan Mo, Zahra Tarkhani, and Hamed Haddadi.
\newblock Sok: Machine learning with confidential computing.
\newblock arXiv, 2022.

\bibitem{flateeFedLearningTEE}
Arup Mondal, Yash More, Ruthu~Hulikal Rooparaghunath, and Debayan Gupta.
\newblock Flatee: Federated learning across trusted execution environments.
\newblock arXiv, 2021.

\bibitem{MlTrainingTee}
Tsunato Nakai, Daisuke Suzuki, and Takeshi Fujino.
\newblock Towards trained model confidentiality and integrity using trusted
  execution environments.
\newblock In Jianying Zhou, Chuadhry~Mujeeb Ahmed, Lejla Batina, Sudipta
  Chattopadhyay, Olga Gadyatskaya, Chenglu Jin, Jingqiang Lin, Eleonora
  Losiouk, Bo~Luo, Suryadipta Majumdar, Mihalis Maniatakos, Daisuke Mashima,
  Weizhi Meng, Stjepan Picek, Masaki Shimaoka, Chunhua Su, and Cong Wang,
  editors, {\em Applied Cryptography and Network Security Workshops}, page
  151–168, Cham, 2021. Springer International Publishing.

\bibitem{privPresMlTee}
Krishna~Giri Narra, Zhifeng Lin, Yongqin Wang, Keshav Balasubramaniam, and
  Murali Annavaram.
\newblock Privacy-preserving inference in machine learning services using
  trusted execution environments.
\newblock arXiv, 2019.

\bibitem{streambox}
Heejin Park, Shuang Zhai, Long Lu, and Felix~Xiaozhu Lin.
\newblock {StreamBox-TZ}: Secure stream analytics at the edge with {TrustZone}.
\newblock In {\em 2019 USENIX Annual Technical Conference (USENIX ATC 19)},
  pages 537--554, Renton, WA, July 2019. USENIX Association.

\bibitem{iioteed}
Sandro Pinto, Tiago Gomes, Jorge Pereira, Jorge Cabral, and Adriano Tavares.
\newblock {IIoTEED}: An enhanced, trusted execution environment for industrial
  {IoT} edge devices.
\newblock {\em IEEE Internet Computing}, 21(1):40--47, 2017.

\bibitem{iftttVsZapier}
Amir Rahmati, Earlence Fernandes, Jaeyeon Jung, and Atul Prakash.
\newblock {IFTTT} vs. {Zapier}: A comparative study of trigger-action
  programming frameworks.
\newblock arXiv, 2017.

\bibitem{teedef}
Mohamed Sabt, Mohammed Achemlal, and Abdelmadjid Bouabdallah.
\newblock Trusted execution environment: What it is, and what it is not.
\newblock In {\em 2015 IEEE Trustcom/BigDataSE/ISPA}, volume~1, pages 57--64,
  2015.

\bibitem{walnut}
Sandy Schoettler, Andrew Thompson, Rakshith Gopalakrishna, and Trinabh Gupta.
\newblock Walnut: A low-trust trigger-action platform.
\newblock arXiv, 2020.

\bibitem{vc3}
Felix Schuster, Manuel Costa, Cédric Fournet, Christos Gkantsidis, Marcus
  Peinado, Gloria Mainar-Ruiz, and Mark Russinovich.
\newblock {VC3: Trustworthy Data Analytics in the Cloud Using SGX}.
\newblock In {\em 2015 IEEE Symposium on Security and Privacy}, pages 38--54,
  2015.

\bibitem{healthcareDataBreach}
Adil~Hussain Seh, Mohammad Zarour, Mamdouh Alenezi, Amal~Krishna Sarkar, Alka
  Agrawal, Rajeev Kumar, and Raees Ahmad~Khan.
\newblock Healthcare data breaches: Insights and implications.
\newblock {\em Healthcare}, 8(2), 2020.

\bibitem{homomorphicPerformance}
Vasily Sidorov, Ethan Yi~Fan Wei, and Wee~Keong Ng.
\newblock Comprehensive performance analysis of homomorphic cryptosystems for
  practical data processing.
\newblock arXiv, 2022.

\bibitem{teeCloudFog}
Dalton Cézane~Gomes Valadares, Newton~Carlos Will, Marco~Aurélio Spohn,
  Danilo~Freire de~Souza~Santos, Angelo Perkusich, and Kyller~Costa Gorgônio.
\newblock Trusted execution environments for cloud/fog-based {Internet} of
  {Things} applications.
\newblock In Markus Helfert, Donald Ferguson, and Claus Pahl, editors, {\em
  Proceedings of the 11th International Conference on Cloud Computing and
  Services Science, CLOSER 2021, Online Streaming, April 28-30, 2021}, pages
  111--121. SCITEPRESS, 2021.

\bibitem{intelSgxSystematicStudy}
Newton~Carlos Will, Dalton~Cézane Gomes~Valadares, Danilo~Freire
  De~Souza~Santos, and Angelo Perkusich.
\newblock Intel software guard extensions in internet of things scenarios: A
  systematic mapping study.
\newblock In {\em 2021 8th International Conference on Future Internet of
  Things and Cloud (FiCloud)}, pages 342--349, 2021.

\bibitem{surveyFedLearning}
Chen Zhang, Yu~Xie, Hang Bai, Bin Yu, Weihong Li, and Yuan Gao.
\newblock A survey on federated learning.
\newblock {\em Knowledge-Based Systems}, 216:106775, 2021.

\bibitem{performanceOfIntelSgx}
ChongChong Zhao, Daniyaer Saifuding, Hongliang Tian, Yong Zhang, and ChunXiao
  Xing.
\newblock On the performance of {Intel SGX}.
\newblock In {\em 2016 13th Web Information Systems and Applications Conference
  (WISA)}, pages 184--187, 2016.

\bibitem{opaque}
Wenting Zheng, Ankur Dave, Jethro~G. Beekman, Raluca~Ada Popa, Joseph~E.
  Gonzalez, and Ion Stoica.
\newblock {Opaque: An Oblivious and Encrypted Distributed Analytics Platform}.
\newblock In {\em 14th USENIX Symposium on Networked Systems Design and
  Implementation (NSDI 17)}, pages 283--298, Boston, MA, March 2017. USENIX
  Association.

\bibitem{beekeeper}
Lijing Zhou, Licheng Wang, Tianyi Ai, and Yiru Sun.
\newblock Beekeeper 2.0: Confidential blockchain-enabled {IoT} system with
  fully homomorphic computation.
\newblock {\em Sensors}, 18(11):3785, Nov 2018.

\end{thebibliography}
\bibliographystyle{plain}
\newpage
\clearpage

\end{document}